# Measurements on MIMO-FRET nano-networks based on Alexa Fluor dyes

Krzysztof Wojcik, Kamil Solarczyk, and Pawel Kulakowski

*Abstract*—**Nano-communication has gained significant attention in the last few years, as a means to establish information transfer between future nano-machines. Comparing with other communication techniques for nano-scale (calcium ions signaling, molecular or catalytic nanomotors, pheromones propagation, bacteria-based communication), the phenomenon called Förster Resonance Energy Transfer (FRET) offers significantly smaller propagation delays and high channel throughput. In this paper, we report our recent experiments on FRET-based nano-networks performed in the Laboratory of Cell Biophysics of the Jagiellonian University, Kraków. We propose to use Alexa Fluor dyes as nano transmitters and receivers, as they enable to create multiple-input multi-output (MIMO) FRET communication channels and thus enhance FRET efficiency. We measure FRET efficiency values, calculate bit error rates for the measured scenarios and extend the calculations to consider a general case of MIMO ($n$,$m$) FRET channels.**

## I. INTRODUCTION

NANO-MACHINES are envisioned to be artificial mechanical devices that rely on nanometer scale components. They are predicted to be basic building blocks for future nano-robots and nano-processors [1-2]. Their existence will create the possibility to manipulate single molecules of the animate and inanimate matter. They promise an enormous spectrum of applications: from biomedical (genetic engineering, health monitoring, medicine delivery) and industrial (material fabrication) to environmental ones (pollution removing, control over nature ecosystems). Because of its low complexity and size, the possible tasks taken by nano-machines are very limited. In order to perform more complicated operations, these devices must cooperate together, working as large well-organized structures, i.e.

constituting networks. Hence, the crucial issue is to enable efficient *communication* between nano-devices.

The current progress in the area of nano-technology spurred the development of its younger sister: nano-communications. A large number of nano- and micro-scale mechanisms was recently analyzed from the communication view-point: calcium ions signaling [3], flagellated bacteria carrying DNA-coded information [4], molecular motors [5-6], pheromones and pollen propagation [7] and others [8]. Most of them, however, guarantee neither a good throughput nor an acceptably low transmission delay.

Here, we consider the phenomenon of Förster Resonance Energy Transfer (FRET) as a means for efficient nano-communication. It is a spectroscopic process in which one molecule transfers its excitation energy non-radiatively over a distance of 1-10 nm to another molecule [9]. The FRET phenomenon has some critical communication advantages. The first one is a very short timescale of the energy transfer that occurs in nanoseconds. Second, the FRET probability depends on the matching between the emission and absorption spectra of the molecules and decreases with the sixth power of the distance between them. Consequently, FRET can be simultaneously used in different parts of a nano-network, like a wireless spectrum band in a cellular network.

Nano-communication via FRET was already discussed in [10-11] where information theory analysis and simulation results for FRET channels were given. In this paper, we present experimental results of FRET-based communication in nano-networks of antibodies (Immunoglobulin G) labeled with Alexa Fluor (AF) fluorescent dyes. Antibodies have the ability to recognize some parts of target molecules and thus they can be thought of as nano-machines (more specifically: nano-sensors), while AF dyes act as their nano-transceivers transmitting and receiving information bits encoded in quanta of excitation energy. We propose to use AF dyes, as they enable *multiple-input multiple-output* (MIMO) FRET communication. We give the measured FRET efficiencies, calculate bit error rates for the measured scenarios and extend the calculations to consider a general case of MIMO ($n$,$m$) FRET channels. The main contribution of this paper can be listed as following:

- a FRET nano-scale MIMO communication technique based on AF dyes is proposed,
- an overview of fluorescent components feasible for FRET communication is given,
- experimental results of FRET efficiencies for AF-based FRET networks are shown,

The manuscript was submitted February 13, 2014. The work was supported in part by the AGH Grant 11.11.230.018. The confocal instrumentation was purchased through EU structural funds, grant number BMZ No. POIG.02.01.00-12-064/08.

K. Wojcik and K. Solarczyk are with the Division of Cell Biophysics, Faculty of Biochemistry, Biophysics and Biotechnology, Jagiellonian University, 7, Gronostajowa St., 30-387 Kraków, Poland. K. Wojcik is also with the Department of Medicine, Jagiellonian University Medical College, Krakow, Poland (e-mails: krzysztof.wojcik@uj.edu.pl, kj.solarczyk@uj.edu.pl).

P. Kulakowski is with the Department of Telecommunications, AGH University of Science and Technology, Al. Mickiewicza 30, 30-059 Krakow, Poland (e-mail: kulakowski@kt.agh.edu.pl).

- bit error rates for MIMO (*n,m*) FRET channels are calculated.

The rest of the paper contains the following sections. In Section II, the FRET theory and the feasibility of FRET multi-hop communication are considered. In Section III, the details are given on constructing the AF-based FRET networks and other possible components (quantum dots, GFP derivatives) are also discussed. Next two sections contain the description of the laboratory methodology and the obtained measurement results. The bit error rate calculations for MIMO (*n,m*) FRET channels are presented and illustrated in Section VI and finally the paper is concluded in Section VII.

## II. FÖRSTER RESONANCE ENERGY TRANSFER

Förster Resonance Energy Transfer[1] (FRET) is an energy process involving two molecules. The first one, called donor, should be in the excited state, e.g. after absorbing a photon. The second molecule is called acceptor and should be in the ground state. The energy is transferred from the donor to the acceptor (usually 1-10 nm) non-radiatively, without an emission of any photon, by the long range dipole-dipole interaction of both molecules (the explanation of FRET assumes that both molecules are dipoles oscillating with similar frequencies that can exchange energy with each other [12]). The rate of the energy transfer and the range of distances over which it can be observed depend mainly on the overlap of the donor emission spectrum and the acceptor absorption spectrum for a given donor-acceptor pair. These relationships were first measured by Förster [9], and later studied by Stryer and colleagues [13].

If the donor molecule is excited not by an external photon, but the energy comes from a chemical reaction, the phenomenon is called Bioluminescence Resonance Energy Transfer (BRET). BRET has some advantages over FRET. First, the excitation of the donor fluorophore in FRET may also lead to the simultaneous excitation of a small part of acceptor fluorophores, which is obviously not desirable. Second, this excitatory light causes photobleaching[2] of the donor. Third, BRET signal to noise ratio is higher than FRET, thus fewer molecules are needed to reach the same signal level.

### A. FRET rate and efficiency

Let's assume first that an excited molecule exists, but there is no other molecule (potential acceptor) nearby. Then, the energy of the molecule is usually released by a photon emission. It can be also dissipated by other non-radiative mechanisms like quenching or decay processes (Fig. 1).

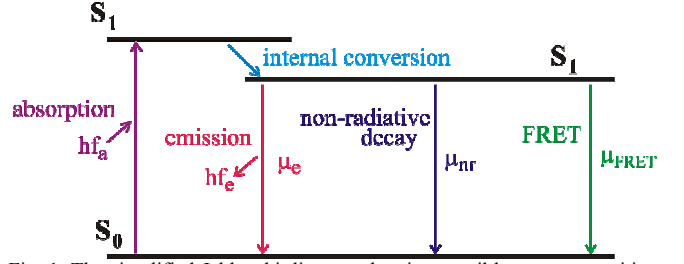

Fig. 1. The simplified Jablonski diagram showing possible energy transitions in an excited molecule.

The time a molecule spends in the excited state before returning to its ground state is called the lifetime of the excited state of the molecule. The average lifetime of the molecule in the absence of FRET transfer can be calculated as the inverse of the sum of its emission rate $\mu_e$ and its rate of non-radiative decay mechanisms (i.e. quenching or non-radiative relaxation) $\mu_{nr}$:

$$\tau_{no-FRET} = \frac{1}{\mu_e + \mu_{nr}} . \tag{1}$$

The emission and the non-radiative decay are random processes with exponential distributions.

It should be noted that some energy is usually lost in the molecule between excitation and emission due to rapid decays between vibrational energy levels (see Fig. 1). The resulting difference between the molecule absorption and emission spectra is called the Stokes shift[3].

Now, if another molecule, being in the ground state, exists in the vicinity of the excited one, the excitation energy can be also released by FRET. In order to have it happen, two conditions must be met. First, both molecules must be in the distance of 1-10 nm from each other. Second, the emission spectrum of the excited molecule, called donor, must overlap the absorption spectrum of the ground state molecule, called acceptor. The FRET rate can be calculated as [12]:

$$\mu_{FRET} = \frac{\mu_e \cdot \kappa^2}{r^6} \cdot \frac{9000 \cdot \ln 10}{128 \cdot \pi^5 \cdot N \cdot n^4} \cdot \int_0^\infty F_D(\lambda) \cdot \varepsilon_A(\lambda) d \cdot \lambda^4 d\lambda \tag{2}$$

The term $\mu_e$ is for the donor emission rate and $r$ is the distance between the molecules. The factor $\kappa^2$ is responsible for the relative orientation between the donor and acceptor dipoles and is usually assumed to be 2/3 which is suitable when both molecules can move by rotational diffusion [12], which is also the case of the experiments reported here. The term $N$ is the Avogadro number, $n$ is the refractive index assumed to be equal to 1.4 for the biomolecules in aqueous solutions. Finally the integral part of the equation is responsible for the molecules spectra overlap: $F_D(\lambda)$ is the normalized emission spectrum of the donor, $\varepsilon_A(\lambda)$ is the absorption spectrum of the acceptor. As the FRET rate depends on the sixth power of $r$, the phenomenon is commonly used in spectroscopy for very accurate measurements of distances between molecules.

---

[1] The abbreviation FRET is sometimes resolved as Fluorescence Resonance Energy Transfer, which however seems to be a misconception, as the phenomenon does require neither a donor nor an acceptor to be fluorescent. Because of that reason, some authors just use the name Resonance Energy Transfer abbreviated as RET. Here, we follow the most common norm, where 'F' in FRET stands for Förster, honoring Theodor Förster (1910-1974), the author of the RET theory.

[2] Photobleaching is a phenomenon of chemical modification of the fluorophore, which in turn loses its ability to emit photons. This is usually caused by breaking open or bonding of the fluorophore to nearby molecules. Although photobleaching is generally assumed to be irreversible, in some cases the fluorophore can be switched on again after loss of emission ability.

[3] For Sir George Gabriel Stokes (1819-1903), who was the first to observe this phenomenon in 1852.

With an acceptor nearby, FRET is another possibility for the donor molecule to dissipate its energy. In general, there can be more that one neighboring molecule in the vicinity of the donor and each of them can be the acceptor in the FRET transfer. All these neighboring molecules contribute to the total FRET rate. Thus, comparing with (1), the donor average lifetime decreases and is equal to:

$$\tau_{FRET} = \frac{1}{\mu_e + \mu_{nr} + \sum_{i=1}^{k} \mu_{FRET,i}} , \qquad (3)$$

where $\mu_{FRET,i}$ is the FRET rate of the $i-$th possible acceptor among $k$ ones. Knowing $\mu_e$, $\mu_{nr}$ and all $\mu_{FRET,i}$ rates, we can also calculate the FRET efficiency $E_i$, i.e. the fraction of the excited donors that decay through FRET to the $i-$th acceptor [12]:

$$E_i = \frac{\mu_{FRET,i}}{\mu_e + \mu_{nr} + \sum_{i=1}^{k} \mu_{FRET,i}} . \qquad (4)$$

The FRET efficiency $E_i$ is also the probability that the energy is transferred to the $i-$th acceptor. Thus, the total probability that the energy is transferred to *any* of the acceptors is given by:

$$E = \frac{\sum_{i=1}^{k} \mu_{FRET,i}}{\mu_e + \mu_{nr} + \sum_{i=1}^{k} \mu_{FRET,i}} . \qquad (5)$$

The distance between the donor and the acceptor where $\mu_{FRET} = \mu_e + \mu_{nr}$ is called the Förster distance $R_0$, which can be measured or calculated with (2), for each donor-acceptor pair. Knowing $R_0$, the FRET rate can be obtained from a simplified formula:

$$\mu_{FRET} = (\mu_e + \mu_{nr}) \cdot (\frac{R_0}{r})^6 . \qquad (6)$$

If there is only one possible acceptor (or others have negligibly small FRET rates), the FRET efficiency can be calculated as:

$$E = \frac{R_0^{\,6}}{r^6 + R_0^{\,6}} , \qquad (7)$$

what means that for $r = R_0$, $E = 50\%$ and $\tau_{FRET} = 1/2 \cdot \tau_{no-FRET}$. If there are many acceptors, the FRET efficiency is given by:

$$E = \frac{R_0^{\,6} \cdot \sum_{i=1}^{k} \frac{1}{r_i^{\,6}}}{1 + R_0^{\,6} \cdot \sum_{i=1}^{k} \frac{1}{r_i^{\,6}}} , \qquad (8)$$

where $r_i$ is the distance from the donor to the $i-$th acceptor. If the acceptors are equally distant from the donor, (8) simplifies to:

$$E = \frac{k \cdot R_0^{\,6}}{r^6 + k \cdot R_0^{\,6}} . \qquad (9)$$

In practice, neither FRET rate nor its efficiency could be measured directly. FRET used to be measured by observing changes in donor's and acceptor's fluorescence intensities. These intensity-based methods, however, have a major drawback due to the influence of photobleaching on measured FRET efficiencies. The alternative, more reliable method of measuring FRET is based on fluorescence lifetimes. What can be observed in this kind of measurements is the decrease of the average lifetime of the donor molecule after the acceptor is added. Having $\tau_{no-FRET}$ and $\tau_{FRET}$ values and applying (1) and (3) to (4), we obtain:

$$E = \frac{\tau_{FRET}^{\,-1} - \tau_{no-FRET}^{\,-1}}{\tau_{FRET}^{\,-1}} = 1 - \frac{\tau_{FRET}}{\tau_{no-FRET}} . \qquad (10)$$

### B. FRET time-scale and throughput

Comparing with different nano-scale communication mechanisms, FRET occurs much faster. The whole process can be divided into three phases. First, a donor molecule absorbs a photon which takes less than a femtosecond ($10^{-15}$ s). Then, the excess of energy is dissipated rapidly by the internal conversion to the lowest vibrational energy level $S_1$, which occurs in about a picosecond ($10^{-12}$ s). After that, the FRET occurs after a random delay equal to the donor excited state lifetime which is of order of nanoseconds. Finally, the donor molecule is again in the ground state, while the acceptor is excited. Summing up the three phases, the delay of the whole energy transfer is of order of nanoseconds. Assuming that a single information bit is going to be sent each time (e.g. '1': the donor is excited, '0': no excitation), the FRET channel throughput can be estimated as tens of Mbits/s.

### C. Multi-hop FRET communication

The FRET phenomenon can occur between any types of spectrally matched molecules, see (2). They can be of biological origin, i.e. organic molecules, or artificial ones, e.g. quantum dots. In life sciences, it is very common to observe FRET for fluorophores, i.e. organic molecules which emission spectra cover the range of visible light. For the purpose of communication via FRET, the molecules emission/absorption spectra can be also located in other parts of the electromagnetic spectrum, e.g. ultraviolet or infrared bands.

It is worth to note that the FRET phenomenon can be used for multi-hop transmissions in molecule chains, where two or more energy transfers occur, one by one. In such a scenario, the acceptor in the first FRET pair is at the same time the donor in the second pair, allowing the signal to pass two or more hops (see Fig. 2a). Because of the Stokes shift, some energy is lost in each hop and the absorption spectrum for the second acceptor lies in the longer wavelengths (lower frequencies) range that for the first acceptor (Fig. 2b). It means that FRET channels are in general one-directional[4]: in Fig. 2a, the transfer A→ B → C is feasible, but C → B → A is very unlikely. As the FRET rate decreases with the sixth power of the distance, it is generally more efficient to transfer a signal via multi-hop FRET than directly. One must assure, however, that all the FRET efficiencies are high enough. The total transfer efficiency is the product of the FRET efficiencies for all the hops. FRET efficiencies in a multi-hop transfer are independent of each other. It is confirmed by the Kasha's rule stating that the emission spectrum of a molecule does not depend on its excitation wavelength [15].

---

[4] There are some exceptions, see the homotransfer phenomenon which occurs e.g. between Bodipy fluorophores [12] or YO-PRO-1 molecules [15].

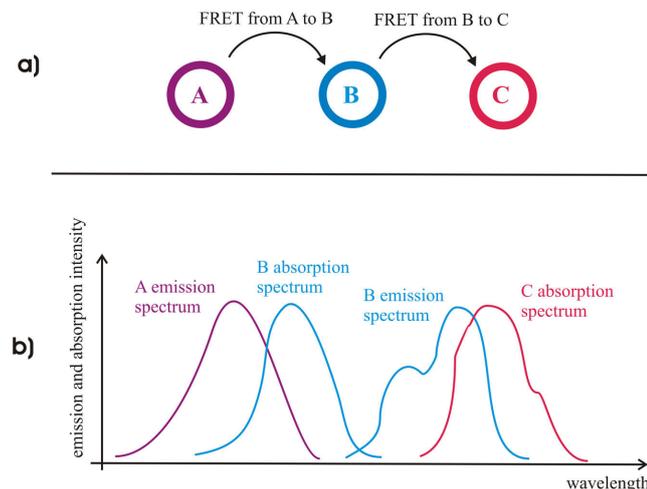

Fig. 2. a) An example of a chain of molecules where multiple FRET reactions can occur. b) Respective emission and absorption spectra of A, B and C molecules.

## III. FRET NETWORKS

The current level of nanotechnology does not allow us to create networks of nano-machines freely moving, approaching each other and communicating. Instead, with the aid of biotechnology techniques, we can construct structures made of proteins, which can potentially work as nano-machines. Certain groups of proteins can perform various nano-machine functions, e.g. (a) kinesins and dyneins are molecular motors able to transport other molecules and (b) antibodies are able to detect specific molecular objects and thus can be treated as nano-sensors. On the top of the proteins, fluorophores can be attached which can communicate via FRET with each other. In general, it means that we cannot create *any* network structure, but we are rather limited by the geometry of the available proteins.

For the purpose of our studies, we built a FRET network composed of 3 antibodies (we used Immunoglobulin G) labeled with fluorescent Alexa Fluor dyes. The Alexa Fluor (AF) dyes are a family of fluorophores characterized by a good robustness against photobleaching. Each fluorophore in the family is usually called Alexa Fluor X, where X is the main wavelength, in nanometers, of its absorption spectrum. The AF dyes can be easily labeled (attached) to antibodies, i.e. proteins of ca. 10-15 nm large. Thus, fluorescent AF dyes can be considered as nano-FRET-transceivers attached to antibodies working as nano-machines (nano-sensors).

While there are other components that can be used in FRET networking (see the summary at the end of this section), we decided to choose the Alexa Fluor dyes mainly for their labeling properties. Using these molecules, each antibody can be labeled not only with a single AF dye, but with a large number of them (theoretically up to 10 dyes), depending on the degree of labeling (DoL). As a result, we had a kind of *multi-input multiple-output* (MIMO) FRET links between the antibodies. As explained in Section VI, having many AF dyes at both sides of each link greatly increased the efficiency of the information transfer.

Our FRET network under study was constructed as follows. The first element of the network was a histone H1 bound to a DNA molecule. Next, a primary antibody Ab0 (ab71594 mouse monoclonal anti-histone H1 IgG) was attached to the histone H1. Then, Ab1 (A31553 goat anti-mouse IgG), Ab2 (A21467 chicken anti-goat IgG) and Ab3 (A11056 donkey anti-goat IgG) antibodies were subsequently added (Ab2 and Ab3 were together attached to Ab1) in order to form the whole FRET network. Each of these 3 antibodies had been already labeled with AF dyes (see Fig. 3).

In general, antibodies allow building practically infinite chains; on the other hand the distance between them cannot be controlled precisely. Our experiments were performed on HeLa 21-4 cells (a common laboratory cell line derived from a human ovarian cancer) containing about $10^7$ of these histone H1 molecules in their nuclei. In the laboratory process of immunostaining, about 30–50% of histones are labeled with antibodies, thus we were able to create about $3–5 \cdot 10^6$ similar fluorophore networks, which we assumed was a good statistical sample.

The main aim of our experiments was to measure the efficiency of the information transfer in all the three communication links created between the antibodies: Ab1-Ab2, Ab2-Ab3 and Ab1-Ab3. Each used antibody had a shape of a 3-element airscrew [16] with a radius of ca. 7 nm (see Fig. 3). The distances between the antibodies Ab1-Ab2 and Ab1-Ab3 were about 9-12 nm, while Ab2-Ab3 distance is smaller, about 7-8 nm. The antibodies Ab1, Ab2 and Ab3 were labeled with Alexa Fluor 405, 488 and 546 dyes, respectively. AF dyes are rather small molecules, about 1.4-1.8 nm large. We used AF dyes produced by Molecular Probes, Inc., having the following degrees of labeling:

- for AF 405, DoL = 2,
- for AF 488, DoL = 7,
- for AF 546, DoL = 4.

Due to chemical properties of antibodies, AF dyes always attach to their upper elements. The exact positions of the dyes on antibodies are, however, hard to determine. The upper element of each antibody is about 7 nm long, the dyes surround these elements. As the result, the distances between the AF dyes may vary significantly. We can estimate them as:

- AF 405 − AF 488 distances = 7-14 nm,
- AF 488 − AF 546 distances = 6-12 nm,
- AF 405 − AF 546 distances = 7-14 nm.

The absorption and emission spectra of the AF dyes are shown in Fig. 4. The Förster distance $R_0$ between AF488 and AF546 is equal to 6.4 nm [17]; the $R_0$ values for AF405→AF488 and AF405→AF546 FRET transfers are, to the best of authors' knowledge, not published. The energy transfers in the opposite directions can be neglected as highly unlikely, what can be confirmed with Fig. 4, as the respective emission and absorption spectra do not match. It means that all the three communication links are unidirectional.

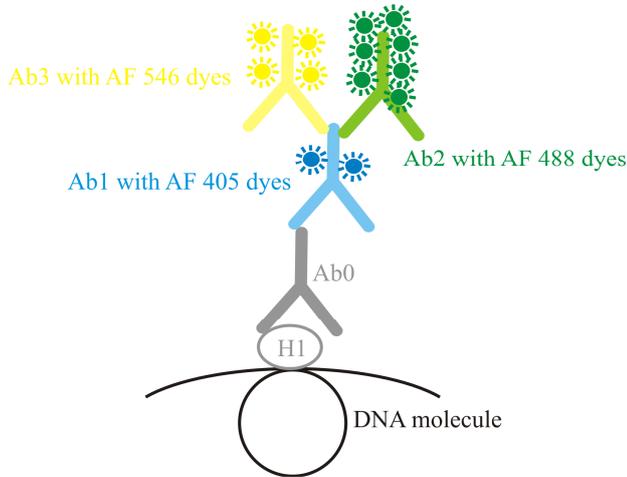

Fig. 3. The network of antibodies (Immunoglobulin G) and AF dyes used in the experiments.

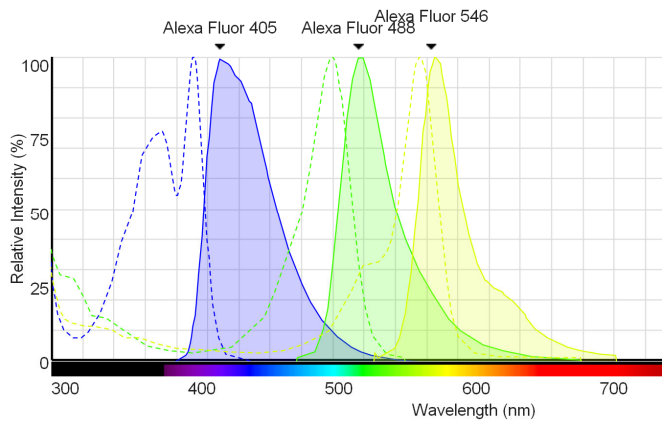

Fig. 4. Absorption (dashed lines) and emission (shaded areas) spectra of Alexa Fluor dyes, by courtesy of Life Technologies [18].

Alexa Fluor dyes are not the only molecules that can be used as nano-transceivers in FRET networks. Other fluorescent components feasible for FRET networking are:

1. A protein family derived from green fluorescent proteins (GFP). A GFP molecule is a barrel shape protein 4 nm high and 3 nm in diameter. The chromophore is hidden inside the structure, which makes it resistant to quenching by external molecules e.g. oxygen. This spatial structure is also responsible for a very good stability of GFP fluorescence lifetime. The proteins derived from GFP (blue, cyan, green, yellow and dsRED) have even better fluorescent properties and can absorb and emit photons in different spectrum bands [19].

Molecular biology techniques create possibilities for building long chains of fluorescent proteins, controlling precisely the distances between the molecules and connecting them with proteins performing various functions. These additional proteins can be used for routing of signal in multi-hop FRET networks. Another class of fluorescent proteins are photoactivatable proteins, which can be turned on by a pulse of light (these molecules become fluorescent proteins after activation by proper light illumination) and make the multi-hop FRET network functional (able to signal transduction). It should be noticed that fluorescent proteins need certain environment to be functional – aqueous solution with proper ion concentration, pH etc.

2. Quantum dots (QDs) nanocrystals are fluorophores made of a semiconductor material. Dimensions of QDs vary from 2.3 nm (blue QD) to 5.5 nm (red QD) in diameter. The core is made of cadmium and the semiconductor shell is usually made of zinc sulfide. The main advantages of QDs are the excellent quantum yield and photostability [20]. QDs are inorganic compounds and in order to integrate them into a multi-hop FRET network these particles need to be attached to other organic structures. This is usually achieved by coating QDs with small organic particles with –COOH chemical group. The –COOH group may form a covalent bond with proteins or organic polymers. Using polymers with regular structure as a backbone allow controlling precisely the distances between QDs. The broad excitation spectrum may be however a disadvantage. Thus, the transmission in multi-hop FRET networks with QDs should be based on bioluminescence phenomena (BRET). QD-BRET is a common technique used in molecular studies [21]. QDs can be used in aqueous as well as in gas environment.

A list of typical Förster distances for different molecule pairs, AF dyes, GFP-derived proteins and quantum dots, is given in Table 1 [17, 22-23].

TABLE I
FÖRSTER DISTANCES FOR DIFFERENT MOLECULE PAIRS

| Alexa Fluor dyes | | GFP derivatives | | Quantum dots | |
|---|---|---|---|---|---|
| Donor/ acceptor | $F_0$ [nm] | Donor/ acceptor | $F_0$ [nm] | Donor/ acceptor | $F_0$ [nm] |
| AF350/488 | 5.0 | EBFP/EGFP | 4.1 | 530 nm QD/Cy3 | 6.0 |
| AF488/546 | 6.4 | EBFP/DsRed | 3.2 | | |
| AF488/555 | 7.0 | ECFP/EGFP | 4.8 | 530 nm QD/Cy5 | 4.9 |
| AF488/594 | 6.0 | SCFP3/SYFP2 | 5.4 | | |
| AF546/568 | 7.0 | mCerulean3/mVenus | 5.7 | 530nmQ D/Cy5.5 | 4.6 |
| AF546/647 | 7.4 | mTurquoise2/mVenus | 5.8 | | |
| AF568/647 | 8.2 | Clover/mRuby2 | 6.3 | 530 nm QD/Cy7 | 3.8 |
| AF594/647 | 8.5 | mKo/mCherry | 6.4 | | |

## IV. LABORATORY METHODOLOGY

### A. Cell culture and immunofluorescence

The HeLa 21-4 cells used in our experiments were maintained in Dulbecco's Modified Eagle Medium (Sigma-Aldrich) supplemented with 10% fetal bovine serum (Gibco, UK) at $37^0$C and 5% $CO_2$. Before adding the primary antibody Ab0 (which binds the histone H1), incubation in 3% bovine serum albumin was performed in order to prevent nonspecific binding [24]. After 1h of incubation with Ab0, cells were washed with phosphate-buffered saline and then, the secondary antibody Ab1 labeled with AF405 was added for the next 1h incubation. Subsequently the antibodies labeled with AF488 and AF546 were added in order to build the whole FRET network.

## B. FLIM measurements

Since it is impossible to record the time decay profile of the signal from a single excitation-emission cycle, fluorescence lifetimes were measured in the Time Correlated Single Photon Counting (TCSPC) mode. This method is based on an accurate measurement of the time between the excitation pulse and the arrival of the first photon to the SPAD (Single Photon Avalanche Diode) detector. Numerous repetitions of this measurement permit to build a histogram, which represents the time decay of the signal. Fitting the resulting histogram with an exponential curve allows extracting the lifetime of a fluorophore. In FLIM, because each pixel of the image contains its own histogram, it is possible to obtain information about the spatial distribution of fluorescent lifetimes in a defined region of the sample material (e.g. a separate cell).

FLIM (Fluorescence Lifetime Imaging Microscopy) measurements were done using a Leica TCS SP5 II SMD confocal system (Leica Microsystems GmbH) integrated with FCS/FLIM TCSPC module from PicoQuant GmbH (Fig. 5). This module consists of picosecond laser diode heads (excitation at 405, 470 or 640 nm), coupled to a laser driver (PDL 828 "Sepia II"), two SPAD detectors (PDM Series), a TCSPC module (PicoHarp 300) and a detector router (PHR 800). In a typical measurement, a FLIM image was acquired at 256x256 pixel resolution and at a speed of 200 lines/s. The acquisition time for each image was one minute, with the laser power set to achieve a photon counting rate of 200-300 kCounts/s (the detection efficiency was about 1-10%). The fluorophores, AF405 and AF488, were excited at a 20 MHz pulse repetition rate with 405 and 470 nm laser lines and the emission was collected with 460-500 and 500-550 band-pass filters, respectively.

It is worth noticing that during each laser pulse, not more than few donor AF dyes were excited among millions of them in the nuclei of HeLa cells. Consequently, we could neglect the scenario with 2 or more excited dyes attached to the same antibody, what is important for the later considerations on MIMO-FRET efficiency (see Section VI).

In each measurement a region of approximately 50x50 μm containing 2-4 cells was chosen. In a given region three lifetime images were collected for AF405: one before and after each staining with secondary antibodies Ab2 and Ab3 (tagged with AF488 and AF546 respectively), while for AF488 two lifetime images were collected: one before and one after staining with AF546-tagged secondary antibody. Before analysis, to exclude the signal resulting from non-specific staining, only the nuclei of cells were chosen. Due to possible differences in the distance between attached antibodies resulting from chromatin compaction, mitotic chromosomes were not analyzed (see Fig. 6 A-F). The analysis was done in SymPhoTime II software (PicoQuant GmbH) using tail fitting, excluding the IRF (Instrument Response Function) time range. To improve signal to noise ratio, all nuclei in a given region were analyzed together, resulting in one decay for each image. Both AF405 and AF488 decays were fitted with a two-exponential function:

$$F(t) = A_1 \cdot e^{-t/\tau_1} + A_2 \cdot e^{-t/\tau_2} , \qquad (11)$$

where $\tau_1$ and $\tau_2$ are the lifetimes, while $A_1$ and $A_2$ represent the amplitudes of the components. The goodness-of-fit was estimated based on the weighted residuals and the chi squared value. For FRET efficiency calculations, see (10), the amplitude-weighted average lifetime was used:

$$\tau = \frac{\sum_i A_i \cdot \tau_i}{\sum_i A_i} . \qquad (12)$$

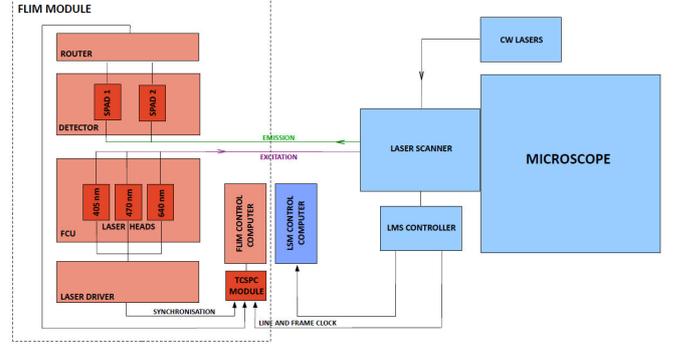

Fig. 5. Schematic representation of the Laser Scanning Microscope (LSM) integrated with a FLIM module. The experiment begins when the sample is excited with a pulsed laser, which is controlled by a laser driver. This unit permits to control the laser power output and its repetition rate. Upon excitation, the sample emits fluorescence photons which are detected in SPAD detectors. In order to determine the exact time between the excitation pulse and arrival of the first photon to the detector, a TCSPC module connected to a computer receives information from the detectors, laser driver and laser scanner. In steady-state measurements (image acquisition without information about fluorescence lifetimes), the sample is excited with a CW (continuous wave) laser, signal is detected in PMT detectors and the resulting image is viewed on a LSM computer.

## C. Image acquisition

The images of the observed cells (see Fig. 6) were obtained with a Leica TCS SP5 II SMD confocal system (Leica Microsystems GmbH), equipped with a 63x NA 1.4 oil immersion lens. Images (8 bit) were acquired at 512x512 pixel resolution, at a speed of 200 lines/s, with 1-3 frames averaged. For AF405 excitation, a 405 nm pulsed laser line was used, while AF488 and 546 were excited with 480 and 543 nm lines (Ar and HeNe lasers). Fluorescence emission was collected using photomultipliers at 460-500, 500-550 and 580-630 nm, respectively. For the comparison of images before and after staining with secondary antibodies, the same laser power and photomultiplier gain was used.

## V. Measurement results

As mentioned in Section II.A, the FRET efficiency cannot be observed directly, i.e. measuring the FRET rate. Instead, the lifetime of the excited state of the donor molecules is compared in two scenarios: without and with the acceptor molecules, see (10).

It should be emphasized that, unlike the FRET efficiency, fluorescence lifetime of a dye depends on pH, ion concentration and presence of other molecules in the solution. Thus, both lifetime measurements, without and with the acceptor molecules, should be performed in the same environment. Having this in mind, we always measured both fluorescent lifetimes on the *same* selected population of molecules: there were 2-4 cell nuclei (see Fig. 6), each contained ca. 3-5 x $10^6$ FRET networks.

FLIM measurements showed the following FRET efficiencies (Table II):

TABLE II
RESULTS OF THE PERFORMED FLIM MEASUREMENTS

| donor | acceptor | $\tau_{no\text{-}FRET}$ [ns] | $\tau_{no\text{-}FRET}$ [ns] | $E$ [%] |
|-------|----------|------------------------------|------------------------------|---------|
| AF 405 | AF 488 | 4.91 | 4.20 | **14** |
| AF 488 | AF 546 | 3.26 | 1.75 | **46** |
| AF 405 | AF 546 | 3.08 | 2.79 | **9** |

Additionally, FRET can be observed visually in confocal images by the decrease of the donor fluorescence intensity in the presence of the acceptor, as shown in Fig. 6.

FLIM measurements and confocal images confirm that the FRET transfer occurred in all three communication links. As we do not know the exact distances between the AF dyes, it is hard to verify if the results of the measurements match the FRET theory. Anyway, we can try to do that for the FRET AF488→AF546, as we know the respective Förster distance: $F_0 = 6.4\,\text{nm}$. In this link, for each donor we have 4 acceptors (for AF 546, DoL = 4). Assuming for the moment that each of these 4 acceptors is located at the same distance $r$ from the donor, we can put the measured FRET efficiency and the Förster distance into the equation (9) and calculate: $r = 8.3\,\text{nm}$. The calculated value matches our expectations: in Section III, the average distance between AF 488 and AF 546 dyes was estimated as 6-12 nm.

## VI. MIMO-FRET CHANNELS

The FRET efficiency measured for all three communication links seems far from being acceptable for telecommunication purposes. Let us adopt, like in [10], the simple ON-OFF keying modulation scheme. We can transmit a single bit '1' – exciting the donor and hoping the energy is transferred via FRET to the acceptor – or a bit '0' – keeping the donor in the ground state, i.e. not exciting it. Assuming no other source of excitation and no other donors in the vicinity, transmission of '0' is always successful and transmission of '1' is erroneous with the probability $1-E$. If zeros and ones are equally probable to be transmitted, the channel bit error rate (BER) is equal to $0.5 \cdot (1-E)$, what means that even in the best measured case (AF488→AF546) the BER is very high: 27%.

Obviously, the channel BER can be decreased by repeating each information bit numerous times, at the cost of the channel capacity. However, a much more efficient solution is to exploit a full multiple-input multiple-output (MIMO) channel, i.e. multiple AF dyes at both sides of each communication link. As stated in Section IV.A, in the performed experiments we had only one donor AF dye excited at the same time. If more donor dyes (attached to the same antibody) are excited, there is larger probability that *at least one of them* transfers the excitation energy via FRET to the donor side. Thus, we can transmit '1' by exciting as many donor dyes as possible. Then, the correct reception takes place when *at least one acceptor dye* receives the energy via FRET. Transmission of '0' is realized as previously, i.e. not exiting any donor. As shown with numbers given below, by using multiple dyes at both sides of the link (MIMO) we can significantly decrease the channel BER in FRET transmissions.

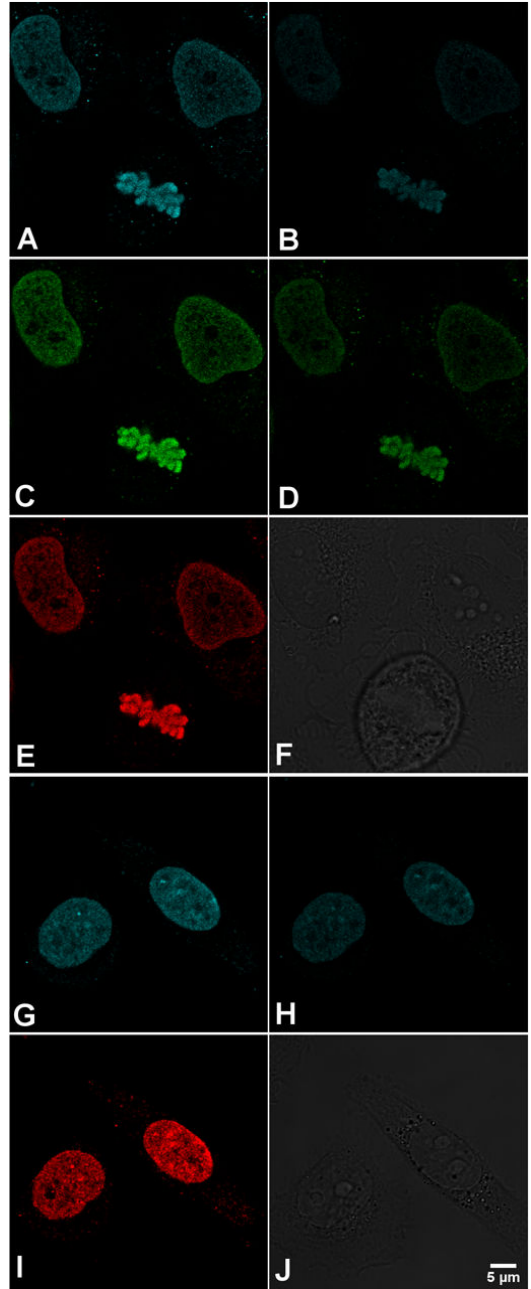

Fig. 6. The decrease of the fluorescence intensity as a result of FRET transfers: AF 405 → AF 488, AF 488 → AF 546 and AF 405 → AF 546. HeLa cells were fixed and incubated with a primary antibody against histone H1. Subsequently, the cells were stained with secondary antibodies (AF 405, 488 and 546) and images were collected before and after staining with each antibody. A) A405 fluorescence before staining with A488. B) A405 fluorescence after staining with A488. C) A488 fluorescence before staining with A546. D) A488 fluorescence after staining with A546. E) A546 fluorescence in cells previously stained with A405 and A488. F) Transmitted light image of the cells showed in A-E. G) A405 fluorescence before staining with A546. H) A405 fluorescence after staining with A546. I) A546 fluorescence in cells previously stained with A405. J) Transmitted light image of the cells showed in G-I.

Let us assume that the FRET efficiency $E$ in the $(1,m)$ communication channel is equal independently which donor AF dye is excited. In fact, what we have measured is the FRET efficiency for different donor AF dyes averaged over tens of millions of excitation events[5]. We can now give the probability of the correct reception of a bit '1' in the $(n,m)$ communication channel which is the probability that *at least one* donor AF dye transfers its excitation energy to an acceptor dye:

$$P = 1 - (1 - E)^n . \qquad (13)$$

Consequently, the bit error rate for the $(n,m)$ channel is equal to:

$$\mathrm{BER}_{n,m} = 0.5 \cdot (1 - E)^n , \qquad (14)$$

where $E$ is the FRET efficiency for $(1,m)$ communication channel measured experimentally or calculated with (5) or (8).

Therefore, we can calculate the $\mathrm{BER}_{n,m}$ values for all three communication links. These are the expected bit error rates in the case when *all* donor AF dyes are excited each time a bit '1' is to be sent through the communication channel. The $\mathrm{BER}_{n,m}$ values are given in Table 3, together with measured FRET efficiencies.

TABLE III
SUMMARY OF THE MEASURED FRET EFFICIENCIES AND CALCULATED BER VALUES

| donor | donor DoL ($n$) | acceptor | acceptor DoL ($m$) | MIMO channel dimension | $E$ for $(1,m)$ channel | BER for $(n,m)$ channel |
|---|---|---|---|---|---|---|
| AF 405 | 2 | AF 488 | 7 | (2,7) | 14 % | **37 %** |
| AF 488 | 7 | AF 546 | 4 | (7,4) | 46 % | **0.7 %** |
| AF 405 | 2 | AF 546 | 4 | (2,4) | 9 % | **41 %** |

We can see that the link AF488→AF546 is clearly the best and the only one with the satisfactory BER value. On the one hand, it is the result of a smaller separation between the donor and acceptor molecules: they are on average 1-2 nm closer than in other two links. Such a difference can be crucial, as the FRET rate decreases with the sixth power of the donor-acceptor separation, see again (2). On the other hand, it is the large number of Alexa Fluor dyes at both donor and acceptor sides (11 in total) that enables to reduce the channel BER. As a reference, we also calculated $\mathrm{BER}_{n,m}$ values for some theoretical cases: donor and acceptor DoL varying from 1 to 10 and donor-acceptor separation from $0.5 R_0$ to $1.7 R_0$ (see Fig. 7). We can see that BER values are increasing rapidly with donor-acceptor distances. The acceptable BER (about 1%) can be obtained even for large distances ($r > 1.5 R_0$), but at the cost of huge number of AF dyes involved. It also seems better to have balanced distribution of dyes at both sides of the link, or a slightly more dyes at donor side than at acceptor side (compare e.g. (4,4), (2,6) and (6,2) cases).

---

[5] The photon counting rate at the detector was at least 200 kCounts/s, each experiment lasted not less than 1 minute and the maximum detector efficiency was 10%, thus we had at least $120 \times 10^6$ photons emitted. The FRET efficiency was at least 6% what means we had at least $127 \times 10^6$ excitation events.

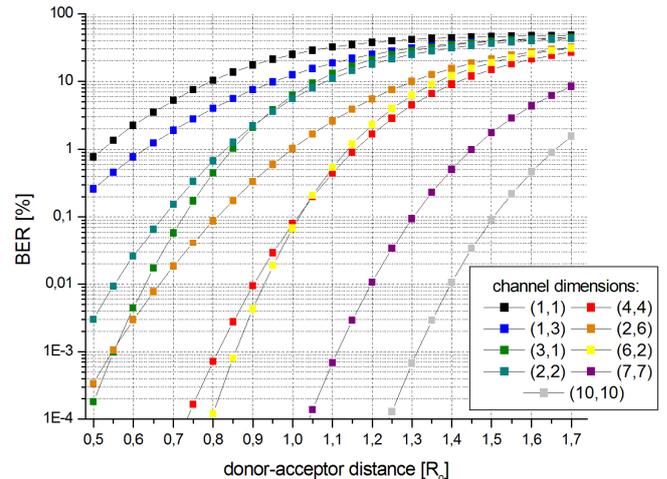

Fig. 7. Theoretical BER values for $(n,m)$ MIMO-FRET channels

## VII. CONCLUSIONS

In this paper, we discuss MIMO communication in nano-networks realized via FRET. We provide an overview of fluorophores feasible for FRET networking. Among them, we propose to use Alexa Fluor dyes as nano transmitters and receivers. These molecules, having degrees of labeling up to 10, can attach in large numbers to proteins working as nano-machines and thus create reliable MIMO-FRET communication channels. We present experimental results of FRET efficiencies for nano-networks of Immunoglobulin G proteins labeled with Alexa Fluor 405, 488 and 546 dyes. We calculate bit error rates for the FRET communication channels under study. Finally, we extend the calculations for a general case of MIMO $(n,m)$ FRET channels. We show that FRET communication can be dependable even on ranges larger than Förster distances if multiple fluorophores are involved at both sides of the channel. Exploiting full MIMO in FRET nano-communication can drastically reduce the channel BER to a level acceptable for reliable data transfers.

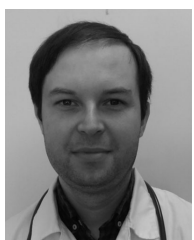

**Krzysztof Wojcik** received the M.Sc. and Ph.D. degrees in biophysics from the Jagiellonian University in Krakow, Poland 2003 and 2015, respectively; and M.D. from the Jagiellonian University Medical College in Kraków, Poland 2007. He was an Assistant at Division of Cell Biophysics Faculty of Biochemistry, Biophysics and Biotechnology Jagiellonian University (2007-2014). He is an Assistant at Allergy and Immunology Clinic in II Chair of Internal Medicine CMUJ. His research interests include the fields of confocal microscopy techniques and its application in autoantibodies research and use of fluorescent labeled antibodies in nanocommunications.

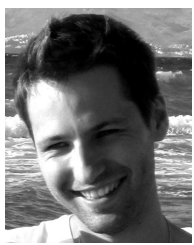

**Kamil Solarczyk** received the M.Sc. degree in biophysics from the Jagiellonian University, Kraków, Poland, in 2010. Since then, he has been working toward the Ph.D. degree in the Faculty of Biochemistry, Biophysics and Biotechnology. His research interests include the DNA repair processes, chromatin architecture and molecular communication.

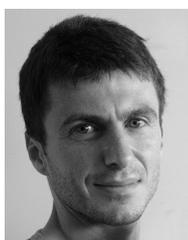

**Pawel Kulakowski** received the Ph.D. degree in telecommunications from the AGH University of Science and Technology in Krakow, Poland, in 2007. Currently he is working there as an Assistant Professor. He was also working few years in Spain, as a visiting post-doc or professor at Technical University of Cartagena, University of Girona, University of Castilla-La Mancha and University of Seville. He co-authored about 30 scientific papers, in journals, conferences and as technical reports. He was also involved in numerous research projects, especially European COST Actions: COST2100 and IC1004, focusing on topics of wireless sensor networks, indoor localization and wireless communications in general. His current research interests include molecular communications and nano-networks. Dr. Kulakowski was recognized with several scientific distinctions, including 3 awards for his conference papers and a scholarship for young outstanding researchers.